\documentclass[a4paper]{article}

\pdfoutput=1

\usepackage{amsmath}
\usepackage{amsfonts}
\usepackage{amssymb}
\usepackage{caption}
\usepackage{subcaption}
\usepackage[pdftex]{graphicx}

\usepackage{color}

\usepackage[square,numbers,sort&compress]{natbib}

\newcommand{\rhostar}{\rho_*}

\title{The effect of magnetic islands on ITG turbulence driven transport}
\author{P. Hill$^1$, F. Hariri$^2$, M. Ottaviani$^1$\\
\small{$^1$CEA, IRFM, F-13108 Saint-Paul-Lez-Durance, France}\\
\small{$^2$EPFL, Lausanne, Switzerland}}

\begin{document}

\maketitle{}

\begin{abstract}

  In this work, we address the question of the influence of magnetic islands on the perpendicular transport due to steady-state ITG turbulence on the energy transport time scale.
  We demonstrate that turbulence can cross the separatrix and enhance the perpendicular transport across magnetic islands.
  As the perpendicular transport in the interior of the island sets the critical island size needed for growth of neoclassical tearing modes, this increased transport leads to a critical island size larger than that predicted from considering collisional conductivities, but smaller than that using anomalous effective conductivities.  

  We find that on Bohm time scales, the turbulence is able to re-establish the temperature gradient across the island for islands widths $w \lesssim \lambda_{turb}$, the turbulence correlation length.
  The reduction in the island flattening is estimated by comparison with simulations retaining only the perpendicular temperature and no turbulence.
  At intermediate island widths, comparable to $\lambda_{turb}$, turbulence is able to maintain finite temperature gradients across the island.

\end{abstract}

\section{Introduction}
\label{sec:intro}

Magnetic islands seriously degrade confinement in tokamaks by flattening plasma pressure profiles.
Understanding the interplay between plasma turbulence and magnetic islands is an important subject of research in magnetized plasmas.
In particular, in the context of the theory of neo-classical tearing modes (NTMs)\cite{Militello2008,Itoh2004,Fitzpatrick1995,Ottaviani2004}
in tokamaks, it is important to assess the competition between parallel and perpendicular transport.

In NTMs, the main drive comes from the lack of bootstrap current in the island region. In the simplest theory, this drive is balanced by a $\Delta'$ term, the tearing mode stability parameter of linear MHD theory, which is the jump of the logarithmic derivative of the perturbed flux function $\tilde{\psi}$ across a rational surface. In the case of NTMs, $\Delta'$ is commonly assumed to be negative, which means that ordinary tearing modes are stable. The size of the island is determined by the balance between the two terms. The bootstrap current drive depends on the degree of pressure flattening in the island region, which in turn depends on the competition between the pressure-flattening parallel transport, and the gradient-restoring perpendicular transport. 

If one ignores perpendicular transport, the theory would give an instability drive inversely proportional to the island width.
For sufficiently small island widths, the bootstrap current drive dominates, so that even when the plasma is stable to the classical tearing mode, it would be unconditionally unstable to the NTM.

The role of perpendicular transport is to curtail the bootstrap current drive so that a critical island width occurs when the effect of parallel and perpendicular transport balance. Below this width, perpendicular transport dominates and the NTM is stable.

If one deals with transport as a diffusive process, the theory of Fitzpatrick~\cite{Fitzpatrick1995} applies. Accordingly, the critical island width $w_c$ for flattening would scale like $w_c \sim (\chi_\perp/\chi_\parallel)^{1/4}$ where $\chi_\perp$ and $\chi_\parallel$ are, respectively, the perpendicular and the parallel transport coefficients of the model anisotropic diffusion problem in the magnetic island geometry.

Within this framework, it would remain to be determined which transport coefficient would be appropriate, whether collisional or turbulent. The two choices would give very different estimates for the critical island size.

However, transport at the scale of an island may not be described by a diffusive process. 
On the one hand, since magnetic islands tend to have vanishing pressure gradient across their width, the region of plasma inside the separatrix, being linearly stable, is expected to be turbulence free.
On the other hand, turbulence excited in unstable regions may penetrate (``spread'') into nearby stable regions.
A key question is therefore to what extent can turbulence propagate across the island separatrix and modify the pressure gradient such that the NTM stability threshold is significantly changed.
In particular, can turbulence alter the critical island width required for the growth of the NTM?
  
Due to the long time scales needed for magnetic islands to evolve to saturation, current research has mainly focused either on 2D two-fluid simulations on transport time scales\cite{Ishizawa2013,Muraglia2011,Ishizawa2009}, or 3D gyro-kinetic/gyro-fluid simulations on much shorter time scales~\cite{Hornsby2012,Poli2009}.

In particular the question of the role of spreading has not been explicitly considered. Addressing this question requires working with a model that has multiple scales. Turbulence must be micro, with a characteristic correlation length of a few ion Larmor radii. This occurs when the model is such that magnetic shear limits the radial size of the turbulent structures as is the case with 3D ITG. Then the island width must be allowed to change from zero to values much larger than the correlation length, in order to explore all possible cases. One must also be able to carry out long simulations reaching overall statistical steady state to measure the various quantities in an unambiguous way. This is achieved when the integration is comparable to the energy confinement time associated with the simulation domain.

In this work, we address this question with direct simulations of a 3D ITG fluid model that satisfy all the above requirements. 
The main result is that the critical island size is indeed determined by the penetration length due to turbulence spreading. This sets a critical island width which would be larger than the one predicted by Fitzpatrick's formula with collisional conductivities but smaller than the predictions of this same formula with anomalous effective conductivities.

The outline of this paper is as follows:
we describe the slab ITG model in section\nobreakspace \ref {sec:model},
the code we used in section\nobreakspace \ref {sec:model} and
the details of the simulations in section\nobreakspace \ref {sec:sims}.
In sections\nobreakspace \ref {sec:width-scan} and\nobreakspace  \ref {sec:discussion} we discuss the results of parameter scans in the island width and the dependence on the normalised gyro-radius, $\rhostar$, and finally discussion and conclusion in section\nobreakspace \ref {sec:conclusion}.

\section{Slab ITG model}
\label{sec:model}

We are interested in understanding the influence of islands on turbulence, so given that the MHD timescales on which the islands grow are much longer than those of the turbulence, we assume that the island is static throughout the simulation.
This work was carried out using a four-field slab electrostatic fluid model similar to the model in \cite{Hariri2014a}, which consists of equations for the normalised density $n$, parallel velocity $u$, and parallel and perpendicular temperatures $T_\Vert, T_\perp$:
\begin{align}
  \partial_t n + [\phi,n] + C_\Vert\nabla_\Vert u &= D_n\nabla_\perp^2 n, \label{eq:dndt}\\
  \partial_t u + [\phi,u] + C_\Vert\nabla_\Vert \{ (1 + \frac{1}{\tau})\phi + T_\Vert \} &= D_u\nabla_\perp^2 n, \label{eq:dudt}\\
  \partial_t T_\Vert + [\phi,T_\Vert] + \frac{1}{L_T}\partial_y\phi + \frac{2}{\tau}C_\Vert\nabla_\Vert u &= \chi_\Vert\nabla_\Vert^2 T_\Vert + D_{T_\Vert}\nabla_\perp^2 T_\Vert,\label{eq:dTpardt}\\
  \partial_t T_\perp + [\phi,T_\perp] + \frac{1}{L_T}\partial_y\phi &= \chi_\Vert\nabla_\Vert^2 T_\perp + D_{T_\perp}\nabla_\perp^2 T_\perp,\label{eq:dTperpdt}
\end{align}
where $\phi$ is the electrostatic potential, $C_\Vert = \tfrac{a}{R}\tfrac{1}{\rho_*}$ with $a,R$ the minor and major radii, $\tau$ is the ratio of electron and ion temperatures, $D_{n,u,T_\Vert,T_\perp}$ are the (artificial, for numerical reasons) perpendicular diffusion coefficients on the respective fields, $\chi_\Vert$ is the parallel heat diffusivity, and $L_T$ is the equilibrium temperature gradient length scale.
We close eqs.\nobreakspace  \textup {(\ref {eq:dndt})} to\nobreakspace  \textup {(\ref {eq:dTperpdt})}  with quasi-neutrality for $\phi$:
\begin{align}
  n = \phi - \rho_*^2\nabla\phi\label{eq:QN}
\end{align}
Note that the usual $\langle\phi\rangle$ term in eq.\nobreakspace \textup {(\ref {eq:QN})} has been removed. This is done for two reasons. On the one hand, in the context of our codes, its implementation would be somewhat computationally burdensome. It is also known that the absence of this term leads to weaker zonal flows than would be expected if it were kept. On the other hand, in the cylindrical model we work with only the slab branch of the ITG mode is present; this branch is weaker than the toroidal ITG branch. In the end the two effects somewhat compensate, yielding a level of turbulent transport in the right range.

Note also that, in our model, the perpendicular temperature is passive; its evolution being determined by Eq.~\ref{eq:dTperpdt} without feeding back into the other model equations.
This feature of passive scalar can be exploited as a diagnostic for the simulated dynamics.

\section{Simulations}
\label{sec:sims}

Our simulation tool is FENICIA, a modular code which can simulate the general class of plasma models:
\begin{equation}
  \label{eq:model-class}
  \partial_t L \cdot S = I \cdot S + E(S),
\end{equation}
where $L,I,E$ are operators on a vector of states $S$.
$L$ and $I$ are time-independent linear operators, whereas $E$ is a nonlinear operator.

The main feature of FENICIA is the technique used for the treatment of the parallel derivative, $\nabla_\Vert$.
The Flux Coordinate Independent (FCI)\cite{Hariri2014,Hariri2013} system is conceptually simple:
the parallel derivative is constructed by tracing the magnetic field lines
from one perpendicular slice to the next, and interpolating to find the
desired quantity. This frees us from using flux coordinates in the
perpendicular plane, and thus allowing complex magnetic geometry
free of singularities.
A more complete discussion of the FCI approach may be found in \cite{Hariri2014,Hariri2013}.

The system is initialised with the same parallel and perpendicular temperature profiles, which are allowed to freely evolve over the course of the simulation.
Since Dirichlet boundary conditions are prescribed at both radial edges, $x_{min},x_{max}$, the average temperature gradient across the box remains constant.
The forcing is then through a ``thermal bath'' with $R/L_T = 5$.
The remaining profiles are initialised to zero, except for small perturbations in the density and the self-consistent perturbations in the potential \emph{via} quasi-neutrality (see eq.\nobreakspace \textup {(\ref {eq:QN})}).

The simulation domain is a 3D sheared slab, periodic in the parallel direction $z$ and the bi-normal direction $y$, with Dirichlet BCs in the radial direction $x$.

We use nominally ``ITER-like'' parameters with an ion gyro-radius $\rho_i = 4$mm.
Taking the middle of the box to be $r=1$m and assuming a minor radius $a=2$m, then $\rhostar = 1/500$.
The box size is $100\rho_i\times 100\rho_i$ in $(x,y)$, which corresponds to $k_y \sim 16$.
Typically, magnetic islands in ITER are likely to have a poloidal mode number of 3--5, so our islands here are somewhat shorter.
This would, however, require a larger number of grid-points in the poloidal direction ($\sim 6000$), and correspondingly larger computer resources.
We use $N_x = N_y = 400$ and $N_z = 32$ grid-points.
The shorter poloidal extent of the island means that the parallel transport will likely be stronger compared to the perpendicular transport than in a realistic island.

The diffusion coefficients are taken to be proportional to $\rhostar$, and $D_{T_\Vert} = D_{T_\perp} = D_u = D_n = 2\times10^{-4}$ for $\rhostar = 1/500$.
$\chi_\Vert$ on the other hand is set to $\rhostar^{-1} = 500$.

The magnetic field is of the form
\begin{equation}
  \mathbf{B} = \nabla \times (\psi \mathbf{\hat{z}}) + \mathbf{\hat{z}}
\end{equation}
with $\psi$ given by
\begin{equation}
  \label{eq:flux}
  \psi(x,y) = -\tfrac{1}{2}(x-x_0)^2 + \hat{b}_x\cos(k_yy),
\end{equation}
where $x_0$ is the location of the rational surface that the island is centred about, $\hat{b}_x$ is the radial magnetic field perturbation which controls the island width, and $k_y$ is the poloidal wavenumber of the island. The island half-width $w$ is given by $(4\hat{b}_x)^{1/2}$.

Due to the relative low cost of these simulations, we are able to run them for 5--6 $\tau_B$ (Bohm times), well into the steady-state phase.
The simulation is started without an island and allowed to evolve until turbulence has developed and saturated ($\sim2\tau_B$), at which point the simulation is stopped.
The simulation is then restarted with an island and run for a further 4 Bohm times.
Assuming gyro-Bohm scaling, the energy confinement time $\tau_E \sim \rho_*^{-1}a^2/\chi_{eff}$ for this simulation box would be $\sim 10-20\tau_B$.

\section{Island width scan}
\label{sec:width-scan}

\begin{figure}[htb]
  \centering
  \includegraphics[width=0.7\linewidth]{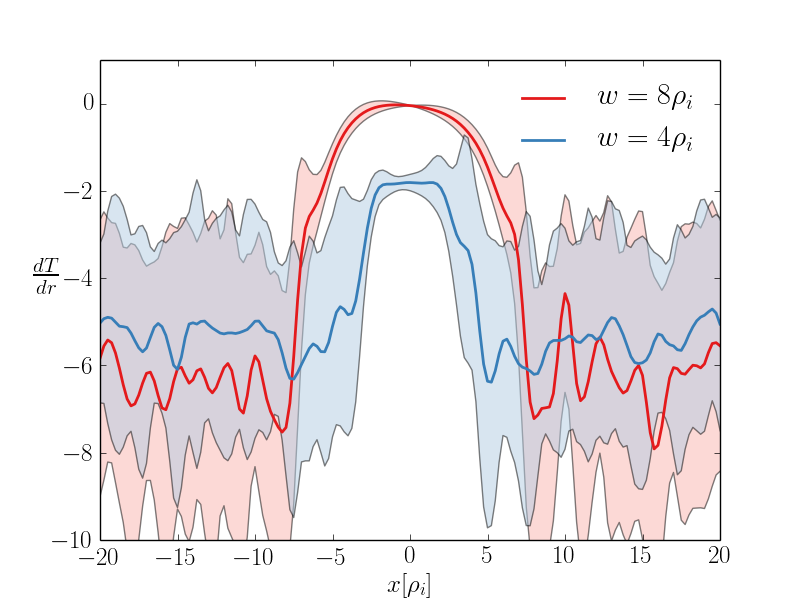}
  \caption{Temperature gradient profile for two island sizes.
    The thick lines are the time-averaged profiles, while the shaded regions are the standard deviations.}
  \label{fig:dTdr}
\end{figure}

Figure\nobreakspace \ref {fig:dTdr} shows the time-averaged temperature gradient profiles along radial chords through the O-points of islands with half-widths $w = 4\rho_i$ and $8\rho_i$.
The standard deviations on the mean profiles give an indication of how far turbulence can penetrate inside the island separatrices.
In both cases, the turbulence can penetrate $\sim4\rho_i$ into the island.
This is roughly equal to the turbulence correlation length $\lambda_{turb} \simeq 3.8$ in the absence of the island.
To measure the correlation length, we first take the cross-correlation of the potential in the radial direction and average over the time and the other spatial coordinates.
An exponential of the form $A + (1-A)\exp[-(\Delta x/\lambda_{turb})^2]$, with $A$ and $\lambda_{turb}$ fitting parameters, is then fitted to the cross-correlation.

\begin{figure}[htb]
  \centering
  \begin{subfigure}{0.5\textwidth}
    \centering
    \includegraphics[width=1.1\textwidth]{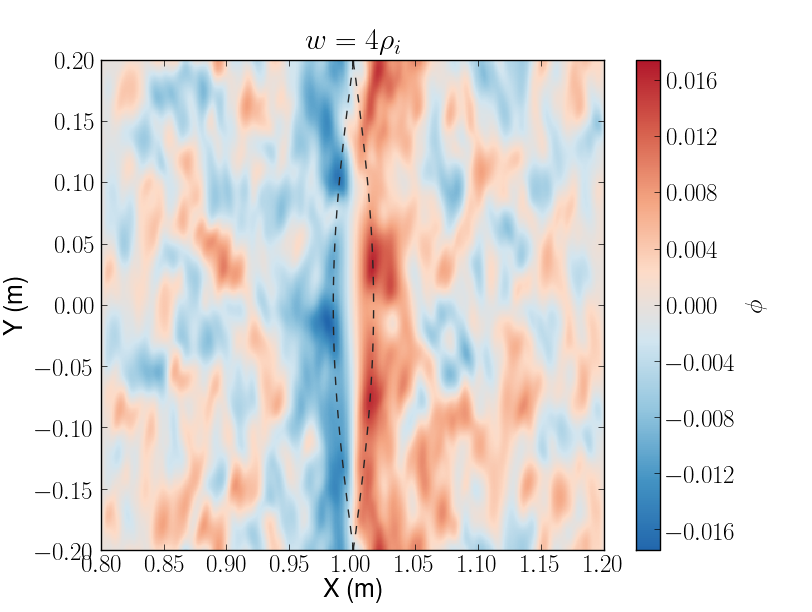}
  \end{subfigure}%
  \begin{subfigure}{0.5\textwidth}
    \centering
    \includegraphics[width=1.1\textwidth]{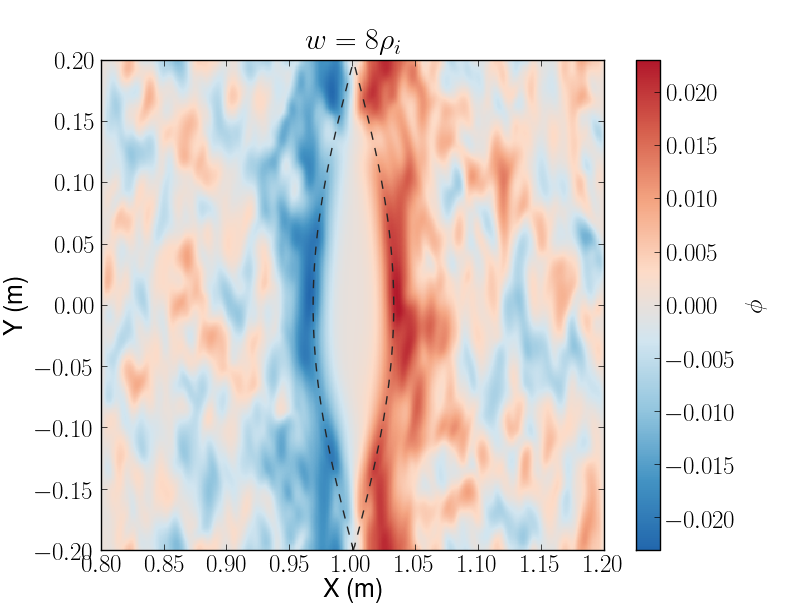}
  \end{subfigure}

  \caption{Snapshots of the electrostatic potential for two different
    island widths: left, $w=4\rho_i$; right, $w=8\rho_i$}
  \label{fig:phi-w4-8}
\end{figure}

This crossing of the separatrix is also visible in fig.\nobreakspace \ref {fig:phi-w4-8}, where the turbulence can be seen to penetrate roughly the same distance into each island.
For islands larger than $\lambda_{turb}$, turbulence is almost completely eliminated from the island centre, whilst for islands smaller than $\lambda_{turb}$, turbulence can still exist at the O-point.

\begin{figure}[htb]
  \centering
  \includegraphics[width=0.7\linewidth]{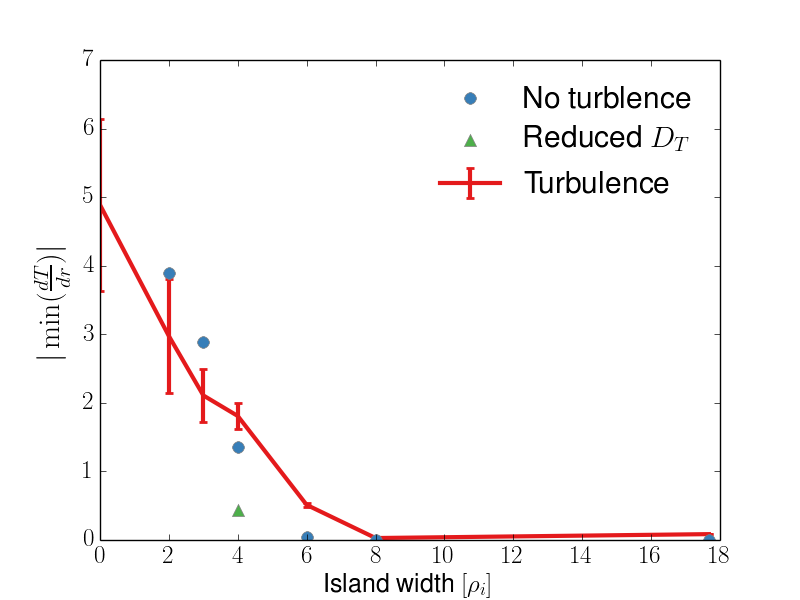}
  \caption{Profile flattening of the perpendicular temperature in the presence of turbulence.}
  \label{fig:f_vs_w}
\end{figure}

To quantify the competition between the parallel and perpendicular transport, we can look at the flattening of the perpendicular temperature profile inside the island.
Use $f=|\min(\tfrac{dT}{dr})|$ 
as a crude proxy for the ratio of parallel and perpendicular transport.
The larger $f$ is, the more perpendicular transport plays a role.
When $f=0$ the temperature profile is completely flattened and parallel transport dominates.
When $f=R/L_T=5$ there is no flattening and the parallel transport can be neglected.
Figure\nobreakspace \ref {fig:f_vs_w} shows how $f$ decreases with increasing island width, with and without turbulence.

Simulations without turbulence were also performed.
This is done by keeping only the equation for $T_\perp$ (eq.\nobreakspace \textup {(\ref {eq:dTperpdt})}), dropping the terms containing $\phi$ but retaining the perpendicular diffusion ($D_{T_\perp}$).

We retrieve the picture from \cite{Fitzpatrick1995}:
in the small island limit, turbulence doesn't see island and so the temperature profile is unaffected by the presence of the island.
Conversely, in the large island limit, the island can (almost) completely expel turbulence and the profile is completely flattened inside the island.
However, we find that for intermediate widths the details of turbulence are important for setting the temperature gradient inside island.
This is most evident for islands close to the large island limit, where turbulence can sustain a finite temperature gradient across the O-point.

The modification of the temperature gradient inside the island due to turbulence has important consequences for the critical seed island width for NTMs.
If turbulence can sustain finite temperature gradients against the parallel transport due to an island, then seed islands will have to be larger than predicted by modelling the perpendicular transport as diffusive.

\begin{figure}[htb]
  \centering
  \includegraphics[width=0.7\linewidth]{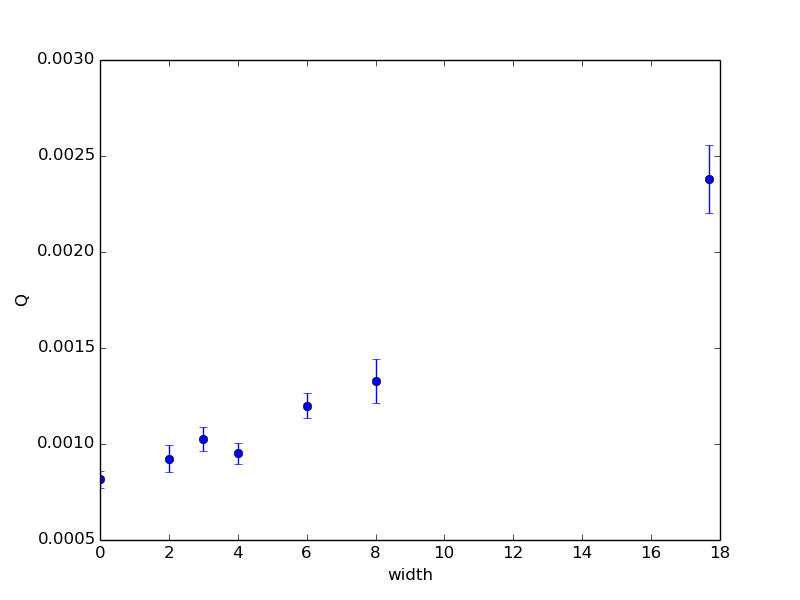}
  \caption{Volume averaged heat flux}
  \label{fig:Q_vs_w}
\end{figure}

Figure\nobreakspace \ref {fig:Q_vs_w} shows the volume- and time-averaged heat flux for increasing island width.
For large islands, heat flux increases with width.
Larger islands have more of an affect on temperature gradient outside separatrix
Island affects temperature gradient immediately outside separatrix, increasing turbulent fluctuations.
There appears to be a threshold around $w=\lambda_{turb}$ below which the heat flux varies very little with island width.
This is in agreement with \cite{Hornsby2010}, despite the different physics models being used, which employs an electromagnetic gyrokinetic model.
There, the threshold also appears at $w\sim\lambda_{turb}\sim 12\rho_i$.

\begin{figure}[htb]
  \centering
  \includegraphics[width=0.7\linewidth]{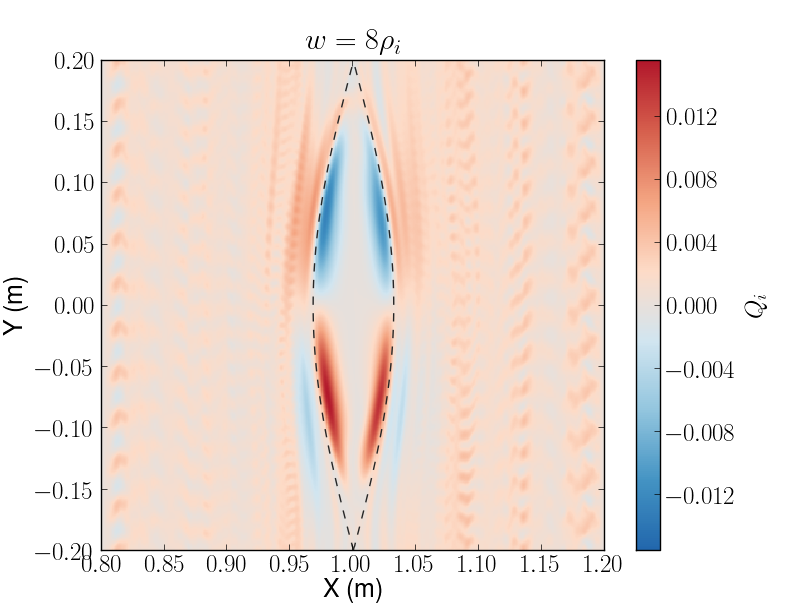}
  \caption{Time averaged poloidal cross-section of the heat flux for an island with $w=8\rho_i$}
  \label{fig:Q_2d_w8}
\end{figure}

For islands above the threshold, the heat flux takes on an up-down asymmetric pattern in the poloidal plane, as illustrated in fig.\nobreakspace \ref {fig:Q_2d_w8} which shows a poloidal cross-section of the heat flux averaged in time and $z$.
The heat flux circulates around the island and most strongly inside the separatrix.

\section{Discussion}
\label{sec:discussion}

Neoclassical tearing modes are a complex phenomenon whose dynamics depends on several elements. One of them is the degree of pressure flattening in the island region, which is the focus of our study.

If transport can be assumed diffusive, the formula given by Fitzpatrick predicts, for the critical island width $w_c$, normalized to the macroscopic scale length,

\begin{equation}
\label{eq:Fitzpatrick}
w_c \sim (\chi_\perp/\chi_\parallel)^{1/4}
\end{equation}

In the collisional regime, the conductivities scale as $\chi_\perp \sim \nu_i \rho_i^2$ and $\chi_\parallel \sim v_i^2 \tau_i$, where the ion collision frequency $\nu_i$ and time $\tau_i$ have been assumed, as it is appropriate for the ion physics under study. 
This collisional estimate gives $w_c \sim (\nu_i/\Omega_{ci})^{1/2}$ which is small for weakly collisional plasmas. 

However, collisional conductivities are not appropriate when the mean-free path is longer than a macroscopic scale length $l$. In this case, one can estimate the parallel conductivity by replacing $\nu_i$ with the transit frequency $v_i/L$ which yields the plateau scaling $\chi_\parallel \sim v_i L$. As for the perpendicular conductivity, one can do the same, which yields $\chi_\perp \sim v_i \rho_i^2 / L$, which is the gyro-Bohm scaling also obtained under the assumption that perpendicular transport is diffusive and ruled by an effective conductivity of gyro-Bohm type, as often assumed in transport modeling.
Putting this together one gets

\begin{equation}
\label{eq:Fitzpatrick_nocoll}
w_c \sim {\rho_{\displaystyle{*}}}^{1/2}
\end{equation}

This is larger than our finding from the previous section, where it is shown that the turbulence spreading mechanism favours a linear dependence on the correlation length. Since the ITG correlation length is proportional to the ion gyroradius one concludes that

\begin{equation}
\label{eq:wc_spreading}
w_c \sim \rho_{\displaystyle{*}}
\end{equation}

One notes that for reactor scale tokamaks such as ITER\cite{Meyer2013,Lawson1957}, which are expected to reach values of $\rho{\displaystyle{*}}$ as low as
$1/500$, the difference between the diffusive scaling~(\ref{eq:Fitzpatrick_nocoll}) and the turbulence spreading scaling~(\ref{eq:wc_spreading}) is substantial.

\section{Conclusion}
\label{sec:conclusion}

Magnetic islands are a serious concern in tokamaks, as they flatten plasma pressure profiles and degrade confinement.
One of the major causes of island formation is the neoclassical tearing mode (NTM), which has been the subject of recent research.
In the context of the theory of NTMs, it is crucial to understand the competition between parallel and perpendicular transport.
There is a critical seed island size, below which the NTMs are stable, which is set by the perpendicular transport.

In this work, we address the question of the influence of magnetic islands on the perpendicular transport due to steady-state ITG turbulence.
This study has shown that turbulence can cross the separatrix of the island, and so sustain a finite temperature gradient across the interior of the island.
The resulting transport therefore changes the critical island width for the onset of NTMs.
The critical island width is larger than that predicted from considering collisional conductivities, but smaller than that from using anomalous effective conductivities.

Turbulence is able to penetrate the island and re-establish the temperature gradient on Bohm time scales.
The saturation of the temperature gradient profile is on turbulent transport timescales,  rather than on collisional ones.
Therefore, one needs to retain turbulence characteristics (e.g. the correlation length) in transport models, rather than modelling the transport as diffusive.

We find that on Bohm time scales, the turbulence is able to re-establish the temperature gradient across the island for islands widths $w \lesssim \lambda_{turb}$, the turbulence correlation length.
  The reduction in the island flattening is estimated by comparison with simulations retaining only the perpendicular temperature and no turbulence\cite{Fitzpatrick1995}.
  At intermediate island widths, comparable to $\lambda_{turb}$, turbulence is able to maintain finite gradients across the island.

This work also shows more evidence for a heat flux threshold in the island width\cite{Hornsby2012}.
Islands with widths below the turbulence correlation length seem to have negligible impact on the heat flux.
Once they grow larger than $\sim\lambda_{turb}$, the heat flux increases linearly with the island width.

The authors would like to thank W. Hornsby and Y. Sarazin for useful discussions.
This work, support by the European Communities under the contract of Association between EURATOM and CEA, was carried out within the framework of the European Fusion Development Agreement.
Part of this work was also supported by the ANR Contract No.E2T2.
Some simulations were performed on the IDRIS supercomputer.

\bibliographystyle{unsrtnat}

\bibliography{./islands_paper}

\end{document}